# Supporting Structured Browsing for Full-Text Scientific Research Reports

Robert B. Allen
School of Information Management
Victoria University of Wellington
Wellington NZ
rba@boballen.info

**ABSTRACT**

Scientific research is highly structured and some of that structure is reflected in research reports. Traditional scientific research reports are yielding to interactive documents which expose their internal structure and are richly linked to other materials. In these changes, there are opportunities to take advantage of the structure in scientific research reports which previously have not been systematically captured. Thus, we explore ways of capturing more of the structure of research in reports about the research and we use that structure to support the development of a new generation of document browsers which include novel interaction widgets. We apply the browsers incorporating the conceptual modeling framework to full-text research reports from the Public Library of Science (PLoS). In addition, we describe the application of model-oriented constructs to facilitating highly interlinked digital libraries.

## 1 INTRODUCTION

Scientific publications are at the dawn of a period of rapid change. We see this in the business models which support them, in their formatting and delivery, in the inclusion of multimedia and simulations, and in the way they are increasingly linked to social media. Traditional scientific research reports are yielding to interactive documents which expose their internal structure and are richly linked to other materials. In these changes, there are opportunities to take advantage of the structure in scientific research reports which previously has not been systematically captured [2].

Research reports may be structured to provide a systematic conceptual overview of the research, especially when those research reports are presented with interactive browsers and coordinated across digital libraries. Thus, we explore ways of structuring research reports to capture more of the rich structure of research and we describe the development of a new generation of document browsers which include novel interaction widgets.

Moreover, we explore tools and issues which may be relevant across many domains. The interlinking of information resources will facilitate new types of interactions. This change will make digital libraries more modular and more like hypertexts or knowledgebases. As such, digital libraries may become more like composite hypertexts [17]. Wikipedia occupies a point in this design space and some of its features may be common in modular digital libraries in the future. For instance, it includes some pages for hypotheses which survey, integrate, and link to research related to a hypothesis. We envision a much richer set of such pages as well as pages for other features such as specific research techniques. Moreover, citations links can be anchored on conceptual models and readers following citations will be able to get context-specific overviews.

In Section 2 of this paper, we discuss basic types of structure for research reports which could be captured. In Section 3 we describe an interactive document browser and some novel coordination widgets implemented with it. In Section 4, we explore the possibilities for causal models in research and in modeling research, and propose a specific framework for them. In Section 5, we elaborate on the conceptual modeling framework and examine how that framework can be applied to example research papers from the full-text open-access collection of the Public Library of Science (PLoS). Section 6 examines organizing research around purposeful action. Section 7 discusses using model-oriented structures to improve the interactive document browser. Section 8 discusses some challenges and considers implications and future work.

## 2 STRUCTURE IN SCIENTIFIC RESEARCH REPORTS

Scientific research reports of a few hundred years ago were generally ad hoc narratives. Over time, they have become increasingly structured. The prevalent type of structure is known as IMRD (Introduction, Method, Results, and Discussion). The general IMRD approach is now so ubiquitous that it is used as the prototypical example of a genre by Swales [27]. Beyond IMRD, recently many research reports are also becoming more structured with widgets such as structured abstracts.

The adoption of text formatting systems such as NROFF and LaTex and mark-up languages such as XML has begun to capture the structure of scientific research reports more systematically. This trend led Harmesze [18] to propose a rich description of structure which could allow for reordering of some report components.

## 3 INTERACTING WITH STRUCTURED RESEARCH REPORTS

The structure of scientific research reports could help improve user interaction both at the level of individual research reports and across a collection of research reports in a digital library.

Early document browsers (e.g., [14]) showed how basic interactive features such as tables of contents and footnotes could be supported. Recent standards such as EPUB support electronic books and browsers using such standards as well as popular consumer products. [6] describes tools which support interactivity in document browsers and electronic books as "coordination widgets". The association of metadata with blocks of text has led to interest in "semantic publishing". Semantic publishing is predicted to change scientific research (e.g., [23]). Indeed [4] went beyond semantic publishing and proposed model-oriented research reports and that text might be largely eliminated from research reports. In the meantime, intermediate steps may incorporate both text and models.

At the same time, digital libraries of traditionally structured research reports have become common. Notable among these is PLoS, which has a large and growing collection of research reports open access and fully reusable with the condition that attribution is made to their source. Each PLoS paper is available in several formats: PDF, HTML, and XML.

Document browsers, semantic publishing, and digital libraries reflect approaches which may be converging but which have not yet been fully integrated. We developed several versions of a browser with widgets for articles from PLoS which achieves some integration of the different approaches. The browser was implemented as a Java applet. While some of the browser's features have been found in other applications, they have not been collected and coordinated in a tool for interacting with collections of scientific documents.

The browser interfaces presented here generally work as described and incorporate the most important features. They are designed for proof of concept rather than as production code that would include refinements such as full line justification and full support for special characters such as Greek letters.

As a test case, we applied the browser program to an article by Zhai et al. [33] which is described more fully in Section 5.2.1. The XML version of this article was obtained from PLoS and a program was developed to segment it based on the XMLSchema tags which are part of its XML file. This component of the program was limited to segmenting articles with followed PLoS markup. As a first step, for displaying the research reports we developed an applet version of the research report with minimal interactivity which approximated the PDF and HTML versions of the reports.

We then added a control panel on the right for navigation to other views and several widgets (Figure 1). One of these widgets is an interactive fisheye table of contents (TOC) similar to the TOC implemented in the generic SuperBook document browser [14]. This TOC collected the first and second level headers which appeared in the document. Those headers were then presented in the lower portion of the control panel. Clicking on one of those TOC labels scrolls the document body text in the center of the screen to the section associated with the label. Unlike the TOC which appears in the HTML versions of PLoS papers, clicking on the applet TOC label also caused subheadings under that label to be displayed.

A second widget developed for this browser added interactivity to the reference list. References are an important part of scientific research reports and recognized as important for semantic publication of scientific research reports. However, while they have been widely studied (e.g., [24, 26, 28]), aside from simple HTML hyperlinks, citations and references have not been much considered in providing interactive services.

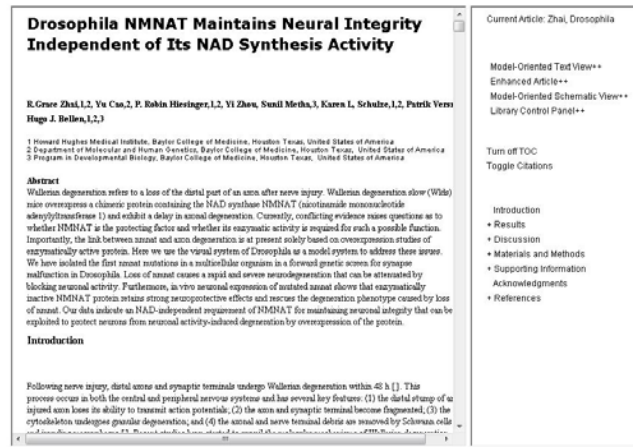

**Figure 1: Applet showing the home page of article with the control panel on the right side. The Table of Contents (TOC) option has been selected and the TOC is shown in the lower portion of the control panel.**

Research publications have differing style guidelines, each with strengths for different tasks. In an interactive research report, styles could be selected by the user. For example, the references and their associated citation numbers could be toggled between being arranged alphabetically or by order of appearance in the text. Our implementation for the interactive Reference List Widget is shown in Figure 2. To accommodate the renumbering of citations in the text, references which indicated multiple citations were broken into separate reference numbers.

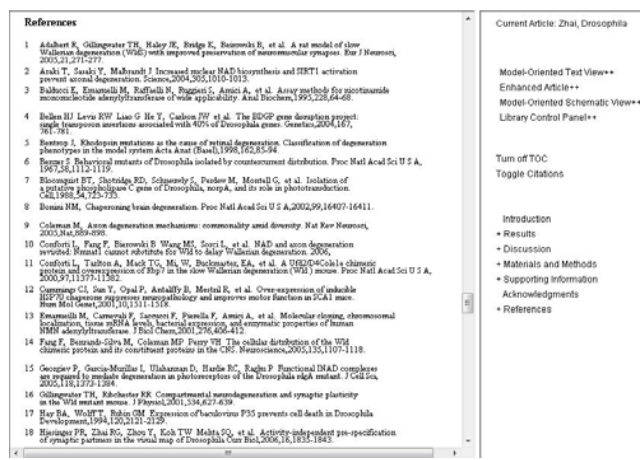

**Figure 2: Interactive Reference List Widget. The control button toggles the presentation between ordering the references in order of appearance and alphabetically by the first author's surname.**

The Reference List Widget is reminiscent of one of the widgets described in [8] which supported lists of archival materials in an interactive finding aid. Two modes for presenting references were implemented in this example. Other modes are possible. For instance, it would be possible to cluster the references by similarity and then to allow exploration of those clusters.

A wide set of additional features could be implemented to enhance interaction with a document. For instance, an interactive list of Figures could be provided. Determining the optimal set of features to include may require human factors testing. Or, many options could be provided and the user could choose among them. We also used the same browser to read in and view Gosby et al. [16] (see Section 5.2.2) although the Tables in that paper were not rendered.

## 4 MODELING THE STRUCTURES OF SCIENCE

The widgets implemented in Section 3 are useful but basic and take little advantage of much of the additional structure in research and research reports. Here, we discuss some of the higher-level structure that may be exploited.

### 4.1 Process Models of Scientific Phenomena

It is widely agreed that models are integral to science while the specifics of those models vary greatly. Some models are structural (e.g., the Bohr atom), others are analogies (e.g., [29]), yet still others are abstract functional relationships. In all cases, identifying causal processes is the critical factor. The importance of causal processes for science has been well recognized (e.g., [19, 32]). Causation can be ambiguous and even probabilistic but it is central to science. Structural models which describe the interacting components of a System, such as the Bohr atom, are validated to the extent they are linked to causal processes and consistent with Occam's razor. Similarly, whether the goal is developing causal abstractions or identifying causes of what happened in a particular situation, causation is a central concern.

Within process models, we emphasize qualitative models rather than functional relationships. Although functional relationships may describe causal processes, we propose that the routine conceptualization of scientific phenomena is most often qualitative. This conclusion is based on observation of how scientists talk and write about their work. For example, it is natural to say that hydrogen and oxygen molecules interact to form water without including a full quantitative description of how that happens. This focus is consistent with the work from cognitive psychology on qualitative reasoning (e.g., [15]). Therefore, while quantitative relationships and feedback loops can be important, at this stage we focused on the qualitative relationships.

### 4.2 Representing Qualitative Models

A further advantage to focusing on qualitative models is the wide range of qualitative modeling tools that have been developed for software engineering and business process engineering. These tools often describe entities which change state as part of some process. [7] developed a state-based approach to causation and we build on that here. The building blocks of that approach are Entities which have States. The State of one Entity is said to be the cause of a State Change (an Event) in another Entity. There are variations and extensions of this basic model. For instance, two Entities may interact to form new Entities as would be the case for the formation of water. Moreover, the models allow the description of Processes and associated Processes at several levels of detail. For instance, we may say that smoking causes cancer but that is clearly short hand for the full details of that Process. The models are developed more fully in Section 5.3 below.

### 4.3 Scientific Research as Identifying Entities and Processes

On the premise that the understanding of causal processes is one of the core goals of scientific research, we propose that incorporating process models into scientific research reports will improve the value of those research reports. For instance, they may allow more focused browsing of the key points or provide links to contextual material for readers who are unfamiliar with the nuances of a field.

Our approach may be distinguished from the studies of scientific research reports which are based primarily on modeling discourse and argumentation (e.g., [12, 13, 25]). Those approaches generally attempt to identify statements about claims and the evidence used to support those claims. By comparison, the present approach asserts that to understand claims the processes about which the claims are being made need to be specified. Indeed, it views the specification of the processes as primary. When not all the component causal processes relating to a phenomenon are understood, the parameters and constraints relating to those unknowns many be specified and then research may be conducted to explore them.

## 5 EXTENDED DESCRIPTION AND LOW-LEVEL IMPLEMENTATION OF CONCEPTUAL MODELING

### 5.1 Approach

To elaborate and demonstrate the potential for the entity-based causal modeling approach we coded causal models in Java (specifically a set of modules within the browser program). This part of the program developed Entity Class and Entity Instance descriptions. Those Entity descriptions were used in Flow descriptions for how they produced new Entities or changes in Entity States. Both Entities and Flows were included in a knowledgebase for which simple relationships were developed. For instance, Entities were related with *part-of* and *type-of* relationships. Similarly, a limited number of types of Flow models were allowed. Given these constraints, we examined the adequacy of the Java-based structural representations and limited extensions of the knowledgebase to reproduce the original research reports.

### 5.2 Example Research Reports for Grounding the Modeling

[4] outlined the application of conceptual models to one of Pasteur's experiments proving germ theory. While seminal, that experiment was simpler than much modern research. In this Section, we describe the application of highly-structured models to two modern research papers that were selected essentially at random from PLoS. We feel that it is essential to consider complete research reports because it helps to ensure

that issues are addressed in context and that the whole spectrum of issues is covered. Indeed, because of their complexity, the two papers turned out to be particularly challenging for implementing conceptual modeling. The paper by Zhai et al. [33] was also used in the description of the browser in Section 3 and is the focus of the analyses here. Gosby et al. [16] provided supplementary evidence.

### 5.2.1 Zhai, et al. [33]

Zhai et al. explores the molecular processes related to the degeneration of axons when they are injured and the processes maintaining their integrity under the stress of normal activity. The research explores the molecular pathways for these processes in *Drosophila*. The research report includes many complex elements. Nearly half of it describes the identification of relevant pathways in *Drosophila*. The report next describes a series of tests that explored using genetic manipulations of those pathways to determine the contribution of various factors. Notably, the activities are generally not described as exploring hypotheses but rather as identifying components of a model.

### 5.2.2 Gosby, et al. [16]

Gosby et al. evaluated the Protein Leverage Hypothesis (PLH), which proposes that the ratio of protein to fats and carbohydrates in the diet affect the number of calories consumed. Ultimately, this may have implications for obesity treatment. Among the goals for Gosby et al. was the evaluation of the PLH in human beings in a randomized clinical trial.

## 5.3 Components

### 5.3.1 Entities Classes and Entity Instances

It is simple to define causal relationships with the model that we have adopted but there needs to be considerable nuance in the structure of each Entity. We employ both Entity Classes and Entity Instances. The Entity Classes may differ in specificity. Thus, there can be a rich linking among the Classes through *part-of* and *type-of* relationships. For instance, both proteins in general and specific types of proteins are Classes but an individual protein that participates in a given reaction is an Instance.

### 5.3.2 Properties and States

Properties are defining values associated with an Entity. The atomic number of an element is an example. By comparison, Attributes are defined as being associated with a given Dimension where the Dimension value is a State. An example is the physical phases of matter: solid, liquid, and gas. The physical phases are dependent on the context of the Entity such as temperature and pressure. The effect of temperature and pressure is specific for each chemical compound and that mapping needs to be associated with it.

We frequently found it necessary to distinguish Entities from collections of those Entities which as a group had a distinct behavior. For instance, a quantity of water such as water in a cup is different from a water molecule. Indeed, the properties of the molecule such as its molecular weight are related to but distinct from the properties of the collection such as its crystal structure.

Thus, while the notion of an Entity is fairly simple, much complexity is incorporated into the details of Entity Properties and States. The specification and management of Properties and States will be central for any knowledgebase developed with this framework.

### 5.3.3 Flows, Relationships, Meshes, and Conceptual Models

*5.3.3.1 Flows*

Flows are process statements about how Entities participate in Events. Specifically, they generally describe discrete-event causal relationship among Entities. They describe some transition due to Entities affecting other Entities. In this paper, we focus on transitions involving qualitative state changes but it should be possible to extend the analyses to causal relationships which are continuous.

There are many factors which may trigger a Flow. For instance a given temperature may be required for a chemical reaction. This constraint could be coded as part of the States of the participating Entities or it could be defined as a Property or Relationship of the Flow itself. The case for defining it as a Relationship is clearer in the case of physical proximity which may be needed before the Event triggers. Flows may be chained together or even be interlaced as Meshes. There are both Class and Instance versions of these Flows. The Classes are high-level generic descriptions while the Instances are specific implementations.

We identify two main types of Flows: Method Flows and Conceptual Models.

*5.3.3.2 Method Flows*

Method Flows are triggered (i.e., caused) by the researcher. They describe Research Methods. Meshes of Method Flows may also be called Workflows.

*5.3.3.3 Conceptual Models*

Some other Flows are Conceptual Models (Figure 3). These are models of explanations for a phenomenon. In some cases, Conceptual Models are competing unproven alternative explanations. In other cases the Conceptual Models are confirmed and accepted explanations for a phenomenon. Conceptual Models may be presented at different levels of abstraction and they may include unknown or under-specified values for Entities and States. Because we view Conceptual Models as abstractions, there are no Instances of Conceptual Models. However, natural phenomena presumably incorporate them [7]. In addition, Conceptual Models may be developed at different levels of detail. In part, this is done by incorporating Entities at different levels of abstraction.

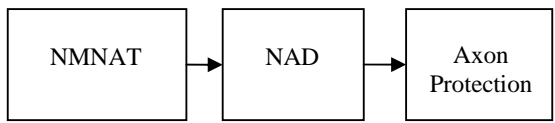

**Figure 3: A simplified Conceptual Model for the effect of the protein NMNAT on the protection of the integrity of axons. A second Conceptual Model did not depend on NAD. One of the experimental manipulations in Zhai et al. disabled NAD production by NMNAT. That State was implemented by creating mutations in the DNA sequences which controlled one of the enzymatic regions of NMNAT. In this model, the State of the NAD is determined by the State of NMNAT.**

### 5.4 Hypotheses and Research Questions

Hypotheses are Conceptual Models to explain a given State Change. Under the proposed approach, Research Questions (RQ) are wrappers around one or more Conceptual Models. White [32] has proposed an ad hoc set of types of Research Questions, but the framework we have developed here provides a more systematic approach. The Research Question may (a) identify some aspects of the Conceptual Models which need to be addressed such as the value of a particular State for a State Change to occur or (b) be the comparison of alternatives. Research Questions can be multi-layered. Just as the details of Flows and Conceptual Models can be refined, Research Questions can be adjusted to be more or less specific. However, Method Flows are not Hypotheses and are not tested with Research Questions.

### 5.5 Systems

Sets of Entities may interact in complex ways. The models described to this point allow for only feed-forward casual relationships involving discrete events. Of course, models with feedback or dynamical systems are possible. We allow for Knots (i.e., feedback loops) in Meshes but we do not develop this point here. It may be that other types of modeling tools such as those from System Dynamics would provide a better foundation for them.

### 5.6 DataSets and Data Recording Instruments

Method Flows describe the Research Method. Some of these Flows are Method Measurement Flows which collect data through instruments. Observations from the instruments are placed into DataSets. DataSets are wrappers which may also include details about the instruments used to obtain and/or record the observations. The DataSets might also include evidence about the reliability and validity (e.g., [9, 10]) of the measurements. In the Java program, we implemented separate class structures (i.e., a construct in the programming language) for DataSets, individual observations (i.e., DataPoints), and data recording instruments.

The research report is useful as documentation to be associated with a DataSet that need to be preserved to understand the context of the data that is collected in the DataSet. Some summary data and statistics will be generated and kept for use in the presentation of the paper. Presumably, full DataSets will be stored to the extent required by editorial policies and legal requirements. It may also be possible to store raw data along with the Workflows and metadata about the statistical packages which are used to generate derived DataSets and statistics.

### 5.7 The Discussion

Among the purposes of the Discussion in a research report are drawing together the research results. This is often the place to address the higher level Research Questions. The Discussion can also provide an examination of the results for related Questions which were not directly addressed in the reported study. The formalism we have developed will be useful for supporting such linking and reconceptualization.

### 5.8 Commentary and Citations

In addition to the purely model-oriented components, a report may include additional Comments. For instance, additional Comments may provide explanations beyond the immediate purposeful action. Presumably they do have discourse purposes which could be identified but we have not yet undertaken a substantial survey of them. Many Comments serve specific discourse functions. A taxonomy of those should be developed and used to enhance the browser. For instance, some Comments may describe the design rationale for choices which were made in conducting the research.

Like Comments, Citations provide additional information about points in the text. Typologies of Citation role relationships such as those developed in [24] are a good place to start. The model-oriented structure can help to provide an unambiguous anchor for those citations.

## 6 DESCRIBING PURPOSEFUL ACTION IN CONDUCTING THE RESEARCH

### 6.1 Activity Blocks

In conducting research, the researcher completes many different actions. In our approach, these actions are described with Method Flows and Meshes. At one level, the sequence of actions moves the presentation forward. The actions are generally accompanied by goals or justifications and also by description of their outcomes. Therefore, typically, the description of scientific activity follows a regular pattern or "Block" in which a goal is stated, a method or set of method of set of methods is applied and the results of those methods are reported. In some cases, a Block is simply aimed at describing the implementation of a method. We term these typical Blocks Activity Blocks (AB).

### 6.2 Research Question Blocks

Research Question Blocks (RQBs) extend the ABs to RQs. Research Questions are the pivots between the Conceptual Models and the Research Methods. In addition to using the RQ as the goal, the RQB should also include an indication about the answer to the RQ. More broadly, the RQBs could also include references to the prior literature as part of the justification and a full discussion of the implications.

Swales's entire IMRD can be viewed as an RQB although IMRD is typically applied to simpler research reports than we are considering here.

Semantic publishing offers the potential for the rearrangement of sections of the text. However, the extent to which sections can be rearranged, especially without a systematic description of the content, remains unknown. The content descriptions of the model-oriented approach should facilitate rearrangement, but even with content descriptions, it is unclear how much flexibility is possible.

### 6.3 Describing Blocks in the Example Papers

In practice, there are several challenges in identifying Activity Blocks. For example, the development of new materials, tools, DataSets and techniques clearly can be constructive scientific activities. In these cases, the researcher's question may be: Can a specific technique be developed for a given application? While no causal model is being evaluated, the answer can be essential for the success of research and thus this activity is an integral aspect of science. Therefore, it seems desirable to identify and differentiate such ABs.

A more difficult distinction arises in the first half of the Results in Zhai et al. [33]. There, the goal is to identify the analog of a protein in *Drosophila* which had already been identified in mice. Further, the genetic basis of that protein and some of its behavior were explored. These earlier findings are useful methodologically later in Zhai et al. [33] but they also elucidate causal processes of the protein behavior in *Drosophila*.

While the remainder of Zhai et al. [33] consists of a sequence of tightly packaged RQBs, Gosby et al. [16] has a single extended method section from which many observations were collected. Those observations were later used to test hypotheses. Thus, sets of Method Flows associated with the RQBs for Gosby were relatively redundant and complex. Rather than being contiguous, they needed to be reconstructed from across the paper.

## 7 USING MODEL-ORIENTED STRUCTURES TO IMPROVE THE BROWSER

### 7.1 Anchoring Citations

The conceptual structures can also provide specific anchors for external citations. For instance, later work by Zhai, [34], extended the previous study and cited a number of different contributions of the earlier work which we explored above. Many of the citations in the later work were to specific procedures and results in the earlier work. Thus, links could be constructed to lead the reader directly to the correct locations in the earlier work. In addition, back-links to more recent publications could be added on the original paper.

When readers shift between papers, especially when they are dropped into the middle of a paper by following a link, they often lack context to fully understand what they are reading. For instance, the link from [34] to [33] refers to neurodegeneration induced by intense light. This effect is introduced on page 2342 of [33] and at a few points later in that paper. However, it is not mentioned at all in the abstract of the paper and appears only indirectly in the table of contents. Thus a reader linking from [34] would probably need some effort to get oriented while attempting to understand what was done.

However, summaries relative to the citation anchor points could be generated to guide a reader. Figure 4 presents an example. While this example was generated manually, it was facilitated by the model analysis and ultimately such summaries may be generated from the model-oriented structural descriptions. Moreover, there are many possible features which could be added to the implementation. Options for personalization should be included and user studies conducted.

### 7.2 Extending the TOCs

As we suggested in Section 3, structure can be the foundation of effective user interaction. Thus, the detailed modeling described in Section 5.3 should be useful for supporting the interactivity of the browser. Following up on the observation in Sections 5.2 and 6.4 about the complexity of the structure in the example papers, we could enrich the original TOC by pointing to the Activity Blocks and Research Question Blocks. The structures of the ABs and RQBs could be linked together for the reader even if they are spread across different sections of the paper. This is similar to the Citation Anchor Widget described in Section 7.1 and an interactive Abstract could also be developed based on the model-oriented structure.

### 7.3 Toward Library-Level Services

Overall, the specification of rich structure within a research report would be useful for many library-level services which link to individual research reports. For instance, a library-level page about the use of a particular research instrument could include a tool to cluster the page's links based on the research methods in which that instrument was employed.

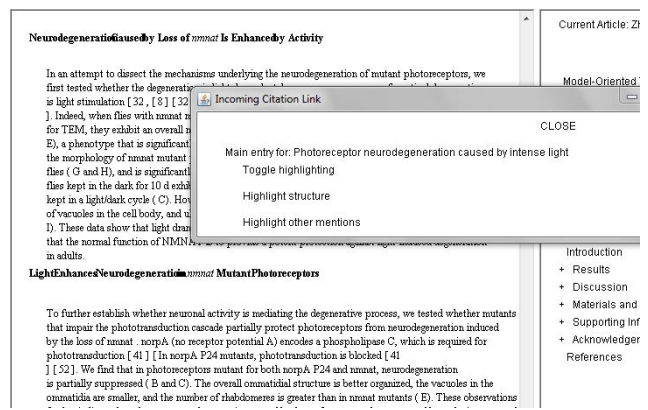

Figure 4: A pop-up frame can provide context for a reader of later publications about a specific anchor point. In this case, the links have been generated manually but potentially they could be derived from model-oriented constructs.

### 7.4 Some Implementation Issues

#### 7.4.1 Incompletely Specified Systems

Developing a general and flexible model for the full complexity of an organism proved particularly challenging. For instance, for Zhai et al. [33] we need to distinguish the representation of a normal *Drosophila* from a *Drosophila* with a mutation. Because we may not know all of the ramifications of that mutation, we must allow for different confidence levels for some attributes. A related challenge is mapping the representation of a mutation in a mouse to an analogous mutation in a *Drosophila*. In the current work, we have coded these higher-level composites (i.e., *Drosophila*, mice) as Systems (cf., [4]). That is, we allowed these cases as Organisms with underspecified elements. The options for the optimal coding need to be explored further, and, the structural framework might be extended to include options for errors, incompleteness, serendipity, and versioning.

### 7.4.2 Managing Inconsistencies

We have adopted a simple Entity and State model to describe the myriad phenomena which science studies. The simplicity of the model is a strength in providing broad coverage and, potentially, in allowing linking across specifications. Nonetheless, there is ambiguity even in relatively simple cases. One example is deciding whether to treat electromagnetic fields as an Entity or a Process. Another example is deciding whether the distal part of an axon relative to a point at which it has been injured should be treated as a distinct Entity. Some of these issues can be managed by editorial policies but eventually they may require the introduction of special cases into the models.

## 8 CONCLUSION

### 8.1 Summary of Major Results

We have proposed a collection of interlocking approaches: a browser for full-text research reports; a framework for developing a knowledgebase of contents for contents of research reports; a new structure for linking the contents of a digital library of scientific research reports; and an approach to formally describing research methods.

We have provided an example and described the potential of browsers for the text of scientific research reports. The browsers take advantage of the various ways in which science and scientific communications are structured. We identify additional useful but currently untapped structure which can be extracted from research reports. We incorporate into the browser program a set of fundamental operators that recreate the elements of each of the research reports studied. We then extend that approach by describing how micro-level conceptual causal models may provide a foundation for structural descriptions. The Java program described here can be viewed as a test bed for continued exploration of how closely the browser widgets can be coordinated with the model-oriented approach.

### 8.2 Extensions

#### 8.2.1 Extending the Browser

Many potential annotation structures are complex. For instance, an external table-of-contents, a timeline, and multi-part argument could include a sequence of points. A multi-anchored annotation widget with sequential constraints among the anchors could be developed to support such structures.

For the immediate future, we envision the enhancements to the browser based on model-oriented structures to improve interaction with text. The design and implementation of these enhancements may be refined through surveys and usability tests with readers. We will also explore a browser for schematic presentations of research reports.

#### 8.2.2 Extending the Middleware

While we have grounded our models in specific examples, these prototypes are modest compared to what would be needed for a large-scale practical system. Several improvements could be readily implemented to the current browser prototype. For instance, the XML files could be preprocessed by a separate program and stored for use. That file could also include the model descriptions.

The modeling resulted in the specification of a very large number of Entities which at times was a challenge to manage. Better authoring support for this process would be useful. Limited support for the automatic extraction of the structure or content of the papers could be useful but generally complex text analysis is so challenging that it will likely continue to be of only limited utility.

The library-level features, which are currently not well developed, could be greatly extended. Search and browsing of the concept hierarchies would be useful at both the browser and library level. In addition, policies and tools should be developed for managing the consistencies of States across research reports.

Overall, this effort might best be approached as an asset of the entire scientific community and the software development might best be an open-source software project.

#### 8.2.3 Challenges and Opportunities for the Management of the Knowledgebase

This work requires developing a knowledgebase. For instance, the *type-of* relationships for Entities described hierarchical linking of Entity types. That knowledgebase is directly linked to rich process models. Some previous efforts to build such knowledgebases have proven impractical. However, the current effort has some advantages over the systems where difficulties were encountered. The domain of this project is scientific knowledge, and scientific knowledge is generally relatively unambiguous. Moreover, we do not envision doing formal reasoning based on the models we have developed. Rather, we expect the models to provide guides human access and browsing. As such, we believe that we can achieve the level of consistency found in the Unified Medical Language System (UMLS) [31]. UMLS has, of course, proven very useful despite some limitations. Ultimately, many of the Entities should be coordinated with domain-specific packages such as the Systems Biology Markup Language (SBML) [20] and the Chemistry Mark-Up Language (CML).

### 8.3 Systematic Descriptions of Research Methods and Designs

While the coding in the browser uses relatively low-level Java constructs, the work here suggests that a more formal set of specifications for activities associated with research can be developed at the level of research methods and designs. We expect that the Method Flow Models which describe the methods applied could be collected and organized into a higher-level language for describing research designs as had been proposed by [4]. We have already described how Conceptual Models may be seen as Hypotheses (Section 5.4) and we have discussed measurements in terms of reliability and validity (Section 5.6).

Swales [27] had described the process of "creating a research space" as one of the key roles of the Introduction in a research report. In a scientific research report Conceptual Models provide the link between the domain and research activity. Taken together, the set of relevant Conceptual Models and Method Flows defines a Research Space. The researcher may not deal with the entire set of possible Conceptual Models but with only a subset of them. This delineation of the Research Space makes clear the extent to which the design of scientific research is a complex process with many tradeoffs. In this sense, Templates for Research Designs [9, 10] may be seen as design patterns.

### 8.4 Implications for the Philosophy of Science

The emphasis on argumentation and discourse structures in scientific research papers seems related to the debates which have come to be known as the "science wars" (e.g., [21, 22]). Some proponents of the discourse approach suggest that science is subjective. The highly-structured and model-oriented approach provides a more traditional positivist view of science. The Conceptual Models are like Hypotheses though they should be seen more as possible alternatives which need to be explored than as ideas which the researcher necessarily prefers. Nonetheless, there may still be Conceptual Models which the researcher prefers and those might be identified as Preferred Hypotheses. Moreover, the model-oriented approach may incorporate argumentation in the way it contrasts conceptual models of prior research.

### 8.5 Causal Models beyond Science

Here, we have emphasized causal models in scientific research reports. There is also a value to linking conceptual and causal models in other areas where discourse and argumentation have been applied, such as design rationale and legal argumentation. [1] described the use of business process workflows as the basis of visualizations for archival records.

We have also explored developing descriptions of the connectedness of events in human history [3]. We can term this an event fabric. There are many similarities in the description of causation for human history with the description of causation for natural phenomena although the development of abstractions for processes is much more challenging.

### 8.6 Envoi

We have developed browsers for text versions of research reports. As predicted in [4], by increasingly taking advantage of the structure inherent in scientific research reports and by linking into standard knowledgebases, the reports may become more schematics than text. While the scope of the current project is already broad, there are still many features which could be added. These include support for collaboration, annotation, personalization, and linking to a broader suite of tools to support the entire range of scientific activities. Ultimately, we envision a library or a set of inter-operating libraries of model-oriented research reports as a complex resource with rich features. This library could be viewed as a Commons with layers of annotations and certifications. Just as there are many editorial boards for paper publications, there could be multiple bodies that certified the contents of these extended features.